# Genetic assignment of illegally trafficked neotropical primates and implications for reintroduction programs.

L. I. Oklander (1), M. Caputo (2), A. Solari (3), D. Corach (2)

((1) Grupo de Investigación en Genética Aplicada (GIGA), Instituto de Biología Subtropical (IBS), Posadas, Misiones, Argentina. Universidad Nacional de Misiones (UNaM) – Consejo Nacional de Investigaciones Científicas y Tecnológicas (CONICET). (2) Departamento de Microbiología, Inmunología, Biotecnología y Genética, Cátedra de Genética Forense y Servicio de Huellas Digitales Genéticas, Facultad de Farmacia y Bioquímica, Universidad de Buenos Aires, Buenos Aires, Argentina; CONICET (3) IBS, Puerto Iguazú, Misiones, Argentina, UNaM – CONICET.)

**Abstract**

The black and gold howler monkey (*Alouatta caraya*) is a neotropical primate that faces the highest capture pressure for illegal trade in Argentina. We evaluate the applicability of genetic assignment tests based on microsatellite genotypic data to accurately assign individuals to their site of origin. The search was conducted on a genetic database to determine the nearest sampled population or to associate them to three clusters described here for the Argentinean populations of *A. caraya*. We correctly assign 73% of the individuals in the database to nearest population of origin, and 93.3% to their cluster of origin. With this database, we were able to determine the probable origin of 17 confiscated individuals, 12 of which were reintroduced in the province of Misiones and 5 confiscated individuals reintroduced in the province of Santa Fe. Moreover, we also determined the probable origin of 3 individuals found dead in cities in northern Argentina. This approach highlights the relevance of generating genotype indexing databases of species to assist with *in-situ* and *ex-situ* conservation and management programs. Our results underscore the importance of knowing the origin of individuals for reintroduction and/or species recovery programs and to pinpoint the hotspots of illegal capture of various species.

**Introduction**

The transformation of natural environments is causing drastic changes in ecosystems at the global level. In Latin America, the development of biodiversity management and conservation plans as part of public policy have not yet taken advantage of newly developed genetic techniques. Conservation genetics can help strengthen the links between scientists and decision makers. Argentina has undergone the replacement of its native forests as a result of the growing urbanization, industrialization and large-scale clearing for agricultural purposes. This process of transforming native subtropical forests is clearly evident in 8 provinces of northern Argentina[1]. Five species of primates inhabit northern Argentina, including *Alouatta caraya, Alouatta guariba clamitans, Aotus azarai, Sapajus nigritus* and *Sapajus cay*. *Alouatta* species are of great epidemiological importance since they are highly sensitive to the Yellow Fever virus (YFV). *A. caraya* show a high mortality rate when infected by the virus. Thus, this species serves as an early epidemiological sentinel, allowing the establishment of control and prevention measures[2,3,4,5].

*Alouatta caraya* is considered "Near Threatened"[6], but is classified as "Vulnerable" in Argentina[7] under International Union for Conservation of Nature (IUCN) criteria "A4cde", due to a reduction in the population, a decreased area of occupancy and/or quality of habitat, increased exploitation due to hunting or illegal traffic (pet trade) and the effects of pathogens and parasites. Because these threats are the same throughout the species' entire range and based on the results of a genetic analysis of Argentinean populations, the global conservation status was recommended to be upgraded to "Vulnerable"[8].

*Alouatta caraya* has been the focus of several population genetics studies focusing on dispersal patterns, kinship, reproductive success and phylogeographic studies, among others[9,10,11,12]. A recent study of the southernmost populations of this species identified different genetic clusters. These data represent an efficient tool that might be used by policy makers when drafting management plans or designing reintroduction projects[8]. One of these clusters or management units (MUs) consists of the populations inhabiting the Atlantic Forest and in the littoral zone bordering Brazil (Misiones cluster[8]). As part of the monitoring program for the epidemiological surveillance of YFV and other arboviroses in non-human primates, the number of genotypes recorded in a database for this area was increased in 2017 and 2018.

Similar to other countries, wildlife trafficking is a threat to Argentinian wildlife. *Alouatta caraya* is the most commonly trafficked primate species in the illegal pet trade in the country. This trafficking is reflected by the increasing frequency of individuals confiscated during control and inspection operations[13]. Official records of the National Wildlife Surveillance and Control-Directorate show that 10 individuals of *A. caraya* have been confiscated and/or voluntarily surrendered in the last 4 years, although the number of unofficial cases is greater. Confiscated and surrendered animals are transported to rehabilitation centres. The return of confiscated animals to the wild (translocation or reintroduction when the translocation occurs inside the species' original range from which it has disappeared) achieves strong support from the public. Although translocations are considered a good option, they might be detrimental for the animal and/or the environment if the policies of returning them to nature are not properly based on scientific evidence.

Species reintroduction has been used to restore top-down trophic interactions and associated trophic cascades to promote self-regulating biodiverse ecosystems in a process called "trophic rewilding"[14]. However, reintroductions can cause undesirable effects, such as the dispersion of invasive species of plants[15]. The reintroductions of animals have multiple risks at the individual, population and ecosystem levels. Therefore, these procedures should undergo a comprehensive risk assessment. The IUCN has published best practice guidelines for adequate reintroductions[16].

The importance of molecular genetics studies enables researchers to identify the geographical region where the individual animals were born (sites of origin). Genetic similarities with a specific previously identified genetic cluster are of the utmost importance. This identification enables researchers to determine the genetic diversity introduced into the environment when translocations and/or reintroductions occur.

The determination of the geographic origin using short tandem repeats (STR; markers that are also known as microsatellites) has already been applied to several species, such as bobcats[17], tortoises[18], African elephants for the ivory trade[19,20], mouflons[21], bears[22], salmon[23], timber[24], palm trees[25] and macaws[26].

In humans, the use of STR-based indexing systems was introduced two decades ago and these systems are the tools used for criminal identification, establishing the identity of fragmented human remains recovered from battlefields, mass disaster environments and other forensic purposes[27]. Moreover, molecular genetic studies allow the identification of species retrieved from the illegal wildlife trade, including the animal products that are traded. A study conducted in South Africa revealed that the products labelled as "game meat" belonged to domestic species in 76.5% of cases[28]. Thus, molecular analyses are helpful for generating a service for both wildlife management and for identification of species to prevent illegal trafficking and prevent consumer fraud.

Genetic assignment to a geographic origin is possible only when a genotype indexing database (GIDB) is available to compare results. The construction of a GIBD represents a requirement and, ideally, should include the largest possible sample across a species' distribution range. Moreover, the data quality must be extremely reliable, because all future determinations rely on the data.

Here, we present the results of genetic analyses designed to determine the origin *A. caraya* confiscated from illegal traders that were housed at the Güira-Oga rescue centre (Güira-Oga) in Puerto Iguazú, Misiones and the Estación Zoológica Experimental Granja la Esmeralda rescue centre (Esmeralda) in Santa Fe, Argentina, and individuals found dead in cities in northern Argentina. We aimed to determine the geographic site of origin of these animals by comparing the specimens' genetic profiles with the profiles included in a regional GIDB presented here for *A. caraya*.

This study underscores the use of STR-based genetic databases to identify the origin of the individuals and provide reliable information to ensure the rigor of translocation/reintroduction program protocols, thus ensuring that they are beneficial for conservation.

**Methods**

*Sampling for GIDB.*

The GIDB for *A. caraya* was built by including the database that was previously complied for many locations in Argentina and contained 143 individuals[8]. The number of samples in the initial database was increased by adding 42 samples obtained from profiles generated from faecal samples collected from several sites in Misiones province (Table 1 and Fig. 1). These samples were obtained during the monitoring program for the epidemiological surveillance of YFV and other arboviroses in non-human primates in 2017 and 2018. Faecal samples were collected non-invasively without capture, and therefore did not cause any harm to the studied individuals. We included 182 individuals in the present study (see the Results for a discussion of repeated individuals). The detection of an identical genotype prompted us to discard one of the samples and three additional individuals were also removed since they shared at least one allele at all loci with other individuals, indicating a first-degree relationship. Accordingly, 178 individuals were included in the overall statistical investigation and four were discarded. Table 1 summarizes the number of individuals, geographical coordinates of sampling locations and type of samples analysed.

*Sampling for genetic assignments.*

Twenty-five samples were collected for genetic assignment. We received hair samples from 17 howler monkeys arriving at Güira-Oga in 2017. Twelve individuals were later reintroduced into a protected area in Isla Palacio at 25°53′32″S, 54°24′38″W (Fig. 1). We also received 5 tissue samples from monkeys arriving at Esmeralda. All these individuals were later reintroduced into a protected area in General Obligado, Santa Fe at 28°00′12.7″S, 59°32′42.09″W (Fig. 1). A detailed description of these individuals is provided in Supplementary Table 1.

Finally, we analysed 3 tissue samples from monkeys found dead by local authorities in Apóstoles, Posadas, and San Antonio, in the Misiones province in northern Argentina (Fig. 1).

*DNA extraction*

All samples were stored at room temperature in 50 ml screw-top tubes containing solid NaCl[29] until DNA extraction (three months to one year later). DNA was extracted from faeces using the QIAamp DNA Stool Mini Kit (QIAGEN, Valencia, USA), according to the manufacturer's protocols with slight modifications. DNA was extracted from tissue and hair samples using standard SDS/Proteinase K digestion followed by phenol:chloroform (1 to 1 volume ratio) organic extraction and Microcon P-100 counter-dialysis filters[30].

*Microsatellite amplification*

Ten microsatellites developed and characterized for *A. caraya* were amplified from each sample: AC14, AC17, AC45, TGMS1, TGMS2, D8S165, D17S804, LL1110, LL1118, and LL157[31,32,33]. Genotyping PCR was performed in a final volume of 25 μl using 5–10 ng of DNA template for tissue samples or 5 μl of the extraction pool from stool samples and included 20 mM Tris–HCl, 50 mM KCl, 1.5 mM $MgCl_2$, 0.2 mM each dNTP, 1 U of GoTaq DNA polymerase (Promega, Madison, USA), 1 pmol of each forward primer bearing an M13 tail, 4 pmol of each reverse primer, and 4 pmol of M13 labelled with a fluorescent dye (6-FAM) on its 5' end using the recommendations from previous studies[31,34]. All amplifications were performed in a Gen Amp ABI 9700 machine (Thermo Fisher, Palo Alto, USA). PCR products labelled with different fluorochromes were combined and the amplicons separated by electrophoresis on an ABI PRISM 310 Genetic Analyzer (Thermo Fisher, Palo Alto, USA). Alleles were manually scored by performing a visual inspection of electropherograms after developing the bin panel for each locus in GeneMapper ID-X v. 1.2 (Thermo Fisher, Palo Alto, USA) using HD400-ROX as internal size standard. For DNA extracted from stool samples, PCR was repeated three times to minimize possible genotyping errors due to allelic dropout[35,36]. We recorded an allele only if it was observed at least twice in different amplifications from the same DNA extract. Homozygous genotypes were replicated three additional times each and scored from the two separate faecal samples per individual. All amplification assays included negative controls.

*Ethics statement*

This study was carried out in strict accordance with Argentinean laws for research on non-human primates, and following the recommendations of 'Principles for the Ethical Treatment of Primates' of the American Society of Primatologists (available at: https://www.asp.org/society/resolutions/EthicalTreatmentOfNonHumanPrimates.cfm). We received specific approval to conduct this study by the Consejo Nacional de Investigaciones Científicas y Técnicas (CONICET) from Argentina (no. 11420110100322CO). Additional specific sampling permits were obtained from Ministry of Ecology, Misiones Province, Argentina (Permit Number: permit no. 9910-00086/17).

*Statistical Analysis*

Genotypes were screened for null-alleles and to discriminate between errors in allele frequency estimates caused by null-alleles, allele dropout or stutter bands using Micro-Checker v2.2.3[37]. Numbers of different alleles, effective and private alleles, observed heterozygosity (Ho), expected heterozygosity (He), unbiased expected heterozygosity (uHe) and inbreeding coefficient were computed with the software GenAlEx v6.5[38] for each locus and population. Deviations from Hardy-Weinberg equilibrium (HWE) were assessed by employing an exact test and FIS:inbreeding coefficient using Arlequin v 3.5 software[39]. Allelic richness was calculated for each locus in a population using the equation $AR = \Sigma \left[ 1 - \left( \frac{\binom{2N-Ni}{2n}}{\binom{2N}{2n}} \right) \right]$, where $Ni$ represents the number of alleles of type $i$ among the $2N$ genes, and n is sample size, using Fstat software v2.9.4[40].

The new complete set of samples collected was analysed using non-spatial Bayesian clustering with the Structure v.2.3.4 program[41]. A series of 20 independent runs per K (ranging from 2 to 6) was conducted using the admixture model with correlated allele frequencies, sampling locations as a prior (LOCPRIOR), and 500,000 Monte Carlo-Markov iterations after a burn-in of 50,000 replicates. The data analysis procedure was further refined using Clumpp software[42] and a bar plot was constructed with the Disrupt software[43]. The most likely number of K was identified using the method described by Evanno[44].

GeneClass2[45] was used to assign the origin of confiscated individuals of unknown origin into the 15 potential populations sampled and the different clusters described here using a leave one-out procedure, excluding self-assignment. After testing all the combinations of approaches presented by the program, we chose the method described by Rannala & Mountain[46], resulting in a higher quality index and the highest number of correctly assigned individuals when tested against the database.

**Results**

Genotypes from populations (Pops) 1 to 10 were obtained in previous studies[8,9]. During the epidemiological surveillance of YFV and other arboviroses in non-human primates in 2017 and 2018, we collected 42 samples and Pops 8 and 9 were resampled. These 42 samples corresponded to 39 different individuals (3 were duplicates, Table 1). All ten loci were amplified for these samples and for the 25 samples for genetic assignment. We did not observe evidence of scoring errors due to stuttering, large allele dropout or null alleles for any loci in any population.

We did not observe evidence of linkage between any pair of loci (P > 0.05). Significant deviation from Hardy-Weinberg equilibrium was only detected for the marker D8a in Pop 9 (Piñalito Province Park). This population had already shown a significant signal attributed to inbreeding in a previous study[8]. The numbers of different alleles, effective and private alleles, observed heterozygosity (Ho), expected heterozygosity (He) and unbiased expected heterozygosity (uHe) are presented in Table 1.

The structure analysis compiling the samples already collected for *A. caraya* in Oklander[8] with the new locations sampled resulted in the definition of three genetic clusters (K = 3, Fig. 2). This new analysis resulted in the disappearance of one previously published cluster (EBCO cluster[8], K = 4, Fig. 2) that now clustered with the populations Chaco National Park, Chaco (Pop 4), Guaycolec, Formosa (Pop 5) and San Alonso, Corrientes (Pop 6, Fig. 2). Accordingly, the 15 locations sampled were grouped into 3 genetically similar regions or clusters. Cluster 1 includes two populations (Pops 1 and 2) from Paraguay-Isla Rio Paraná (P-RP); cluster 2 includes four populations (Pops 3 to 6) from Formosa-Chaco-Corrientes (F-Ch-C); and finally, cluster 3 includes nine populations (Pops 7 to 15) from Misiones-Rio Uruguay (M-RU).

The GeneClass2 software correctly assigned 73% of individuals in the database (Quality index 68.73%) when separated according to the 15 populations and 93.3% (Quality index 89.23%) when separated according to the three clusters. We assigned individuals in the database in groups segregated according to both populations and clusters (Table 2).

The results of the STR-based analysis designed to assess the potential origin of confiscated animals showed that 13 of the 25 individuals assigned showed values less than 75% for the first rank and 6 showed values less than 50% for the first rank (Table 2). This finding was expected because not all populations of *A. caraya* are represented in the database. Therefore, the populations to which the individuals were assigned were interpreted as the nearest sampling site to their real geographic origin. According to the population assignment, of the 17 confiscated individuals housed in Güira-Oga, eight most likely came from a population near Pop 3, three from a population near Pop 2, two from population near Pop 5 and one from populations near Pops 1, 6, 7 and 13, respectively. Of the five confiscated individuals housed in Esmeralda, three most likely came from a population near Pop 3, one from a population near Pop 6 and one from a population near Pop 10. The assignment of the 3 corpses found in cities indicated that the animals most likely came from populations near Pops 5, 12 and 15 respectively (Table 2).

The assignment analysis stratified by clusters revealed values less than 75% for the first rank in only 3 of the 25 individuals assigned and no values less than 50% for the first rank (Table 2). Of the 17 confiscated animals housed in Güira-Oga, only 2 presented a probable origin in the M-RU cluster, while 11 most likely came from the F-Ch-C cluster and 4 from the P-RP cluster (Table 2). Of the 5 confiscated animals housed in Esmeralda, 4 most likely came from the F-Ch-C cluster and only one from the M-RU cluster. The confiscation sites of these 5 individuals were registered, and the animals that clustered in F-Ch-C (individuals Esmeralda 2, 3 and 4, Supplementary Table 1) were captured in the Santa Fe province, while the other two (Esmeralda 1 and 5) were captured in provinces that do not belong to the natural distribution of *A. caraya* (Supplementary Table 1). Of the 3 corpses found in cities, two belonged to the same cluster of the area of the city where they were found, and one belonged to a different cluster (F-Ch-C, Table 2).

In summary of all confiscated individuals (22), 15 individuals and one of the corpses belonged to the F-Ch-C cluster. Therefore, the largest number of illegally trafficked *A. caraya* originated in this area.

**Discussion**

As the first application of the GIDB of howler monkeys, our results indicate that the most likely origins of most of the confiscated and surrendered individuals were populations in the areas of the Pops 2 and 3 (Fig. 1).

This area is also the location of the larger cities in Northeast Argentina and the National highway 12, the main highway connecting these cities with the capital, Buenos Aires. The illegal sale of *A. caraya* has been reported at several locations along this highway[47,48]. This information supports a possible animal trafficking route that begins in northeastern Argentina and ends in Buenos Aires, were the majority of confiscations occur (10 of 22, Supl. Table 1[47,48]). Importantly, most of the confiscations and surrenders occurred in cities outside the normal distribution of the species (17 of 22, Supl. Table 1), indicating that these animals are not only opportunistically captured by locals, but that these animals are intentionally transferred to urban centres as pets. This example illustrates how a genetic analysis helps trace wildlife trafficking routes and hotspots and thus aids in the planning and implementation of more effective control measures.

On the other hand, 15 of 17 animals that arrived at Güira-Oga were assigned to the either to F-Ch-C or to P-RP clusters; twelve of these individuals (confiscated Güira-Oga 1 to 12) were reintroduced near this rescue centre on Isla Palacio, where the local genetic variation belongs to the M-RU cluster, thus introducing genetic variation from animals belonging to different genetic clusters. The 5 individuals that arrived at Esmeralda were also reintroduced in a protected area of General Obligado, Santa Fe (Fig. 1). Although nearby populations are not sampled in the database, we would expect that of our sample areas, genetic variation in the liberation area would be most similar to F-Ch-C, similar to the southernmost area of the distribution of *A. caraya*. Of the five reintroduced individuals, 4 belonged to the same cluster and only one belonged to the M-RU cluster; therefore, these animals have also reintroduced foreign variability, albeit at a lower proportion.

This finding highlights the importance of conducting genetic studies prior to the liberation of rescued animals.

This result also raises the concern of establishing rehabilitation centres in all regions within the three described clusters that could be considered as management units for *A. caraya* if the goal is to reintroduce animals to their native populations.

Conservation genetics is generally not yet well integrated with other efforts in conservation policies. In Latin America, the practical application of genetic principles for the management of threatened species and in the development and implementation of conservation plans should be emphasized. One possible explanation for this disconnect may be that knowledge obtained from scientific research is often not communicated effectively to the field practitioners and/or individuals in power who formulate and enact policies.

As shown in the present study, concrete and measurable genetic data represent a very effective tool to help establish and enforce adequate legislation to curb the loss of biodiversity, generate conservation guidelines, and develop population management strategies that include translocation/reintroduction projects.

## Acknowledgments

We greatly appreciate the collaboration with the Ministerio de Ecología y Recursos Sustentables de la Provincia de Misiones, specifically Patricia Sandoval and Cristina Buhler, and the rescue centers Güira-Oga and Estación Zoológica Experimental Granja la Esmeralda, particularly Jorge Anfuso, Agustin Anzoátegui, Rocio Rodriguez, Antonio Sciabarrasi Bagilet and Pablo Siroski. We thank Silvana Peker and Ricardo Negreira from Secretaría de Gobierno de Ambiente y Desarrollo Sustentable de la Nación Argentina. We are grateful to Sam Shanee for providing comments on earlier drafts of the manuscript. This study was supported by CONICET grants to LO and by DNA Fingerprinting Service (SHDG), School of Pharmacy and Biochemistry, University of Buenos Aires, Argentina. LO, MC and DC are members of the Carrera de Investigator (CONICET-Argentina).

# References

1. UMSEF. 2014. Monitoreo de la superficie de bosque nativo de la República Argentina. Período 2011-2013. Regiones forestales Parque Chaqueño, Yungas, Selva Paranaense y Espinal. (Secretaría de Ambiente y Desarrollo Sustentable, Unidad de Manejo del Sistema de Evaluación Forestal, 2014).

2. Strode, G. K. *Yellow Fever* (McGraw-Hill, 1951).

3. Butcher, E. C. Leukocyte-endothelial cell recognition: Three (or more) steps to specificity and diversity. *Cell* **67**, 1033-1036 (1991).

4. Rifakis, P. M., Benitez, J. A., Rodriguez-Morales, A. J., Dickson, S. M. & de la Paz-Pineda, J. Ecoepidemiological and social factors related to rabies incidence in Venezuela during 2002-2004. *IJBS* **2**, 1–6 (2006).

5. Almeida, M. A. B. *et al.* Surveillance for Yellow Fever Virus in Non-Human Primates in Southern Brazil, 2001–2011: A Tool for Prioritizing Human Populations for Vaccination. *PLoS NTD* **8**, e2741 (2014).

6. IUCN SSC PSG. 2015. Red Listing Workshop for Neotropical Primates. (IUCN SSC Primate Specialist Group, 2015).

7. Oklander, L. I. *et al. Alouatta caraya.* In: Categorización del estado de conservación de los mamíferos de Argentina 2019 in *Lista Roja de los mamíferos de Argentina* (eds SAREM) (SAYDS – SAREM, 2019).

8. Oklander, L. I., Miño, C. I., Fernández, G., Caputo, M. & Corach, D. Genetic structure in the southernmost populations of black-and-gold howler monkeys (*Alouatta caraya*) and its conservation implications. *PLOS ONE* **12**, e0185867 (2017.).

9. Oklander, L. I., Kowalewski, M. M. & Corach, D. Genetic consequences of habitat fragmentation in black-and-gold howler (*Alouatta caraya*) population from Northern Argentina. *Int. J. Primatol.* **31**, 813–832 (2010).

10. Oklander, L. I., Peker, S. & Kowalewski, M. M. The situation of field primatology in Argentina: recent studies, status and priorities. *in A Primatologia no Brasil* (eds Miranda J. M. D. & Hirano Z. M. B.) 31–50 (UFPR/SBPr, 2011).

11. Oklander, L. & Corach, D. Kinship and dispersal patterns in *Alouatta caraya* inhabiting continuous and fragmented habitats of Argentina *in Primates in fragments: complexity and resilience* (eds Marsh L. K. & Chapmans C. A.) 399–412 (Springer, 2013).

12. Oklander, L. I., Kowalewski, M. M. &Corach, D. Male reproductive strategies in black and gold howler monkeys (*Alouatta caraya*). *Am. J. Primatol.* **76**, 43–55 (2014).

13. Bertonatti, C. El comercio de primates en la República Argentina. *Neotrop. Primates* **3**, 35-37 (1995).

14. Svenning, J. C. *et al.* Science for a wilder Anthropocene: synthesis and future directions for trophic rewilding research. *Proc. Natl. Acad. Sci.* **113**, 898–906 (2016).

15. Polak, T. & Saltz, D. Reintroduction as an ecosystem restoration technique. *Conserv. Biol.* **25**, 424–424 (2011).

16. IUCN. Guidelines for Reintroductions and Other Conservation Translocations. Prepinted at https://portals.iucn.org/library/efiles/documents/2013-009.pdf (2013).

17. Millions, D. G. & Swanson, B. J. An application of Manel's model: detecting bobcat poaching in Michigan. *Wildl. Soc. Bull.* **34**, 150–155 (2006).

18. Schwartz, T. S. & Karl, S. A. Population genetic assignment of confiscated gopher tortoises. *J. Wildl. Manag.* **72**, 254-259 (2008).

19. Wasser, S. K. Assigning African elephant DNA to geographic region of origin: applications to the ivory trade. *Proc. Natl. Acad. Sci. U. S. A.* **101**, 14847–14852 (2004).

20. Wasser, S.K. *et al.* Combating the illegal trade in African elephant ivory with DNA forensics. *Conserv. Biol.* **22**, 1065–1071 (2008).

21. Lorenzini, R., Cabras, P., Fanelli, R. & Carboni, G. L. Wildlife molecular forensics: identification of the Sardinian mouflon using STR profiling and the Bayesian assignment test. *Forensic Sci. Int. Genet.* **5**, 345–349 (2011).

22. Andreassen, R. *et al.* A forensic DNA profiling system for Northern European brown bears (*Ursus arctos*), *Forensic Sci. Int. Genet.* **6,** 798–809 (2012).

23. Glover, K. A., Hansen, M. M. & Skaala, O. Identifying the source of farmed escaped Atlantic salmon (*Salmo salar*): bayesian clustering analysis increases accuracy of assignment. *Aquaculture* **290**, 37–46 (2009).

24. Degen, B. *et al.* Verifying the geographic origin of mahogany (*Swietenia macrophylla*, King) with DNA-fingerprints. *Forensic Sci. Int. Genet.* **7**, 55–62 (2013).

25. Nazareno, A. G. & dos Reis, M. S. Where did they come from: genetic diversity and forensic investigation of the threatened palm species *Butia eriospatha*, *Conserv. Genet.* **15**, 441–452 (2014).

26. Presti, F. T., Guedes, N. M. R., Antas, P. T. Z. & Miyaki, C. Y. Population Genetic Structure in Hyacinth Macaws (*Anodorhynchus hyacinthinus*) and Identification of the Probable Origin of Confiscated Individuals. *J. Heredity* **106**, 491–502 (2015).

27. Butler, J. M. & Butler, J. M. *Fundamentals of forensic DNA typing*. (Academic Press/Elsevier, 2010).

28. D'Amato, M. E., Alechine, E., Cloete, K. W., Davison, S. & Corach, D. Where is the game? Wild meat products authentication in South Africa: a case study. *Investig. Genet.* **4**, 6 (2013).

29. Oklander, L. I., Marino, M., Zunino, G. E. & Corach, D. Preservation and extraction of DNA from feces in howler monkeys (*Alouatta caraya*). *Neotr. Primates* **12**, 59-63 (2004).

30. Green, M. R. & Sambrook, J. *Molecular Cloning: A Laboratory Manual, 4th ed.* (Cold Spring Harbor Laboratory Press, 2012).

31. Oklander, L. I., Zunino, G. E., Di Fiore, A. & Corach, D. Isolation, characterization and evaluation of 11 autosomal STRs suitable for population studies in black and gold howler monkeys *Alouatta caraya*. *Mol. Ecol. Notes* **7,** 117–120 (2007).

32. Di Fiore, A. & Fleischer, R. C. Microsatellite markers for woolly monkeys (*Lagothrix lagotricha*) and their amplification in other New World primates (Primates: Platyrrhini). *Mol. Ecol. Notes* **4**, 246-249 (2004).

33. Tomer, Y., Greenberg, D. A., Concepcion, E., Ban, Y. & Davies, T. F. Thyroglobulin is a thyroid specific gene for the familial autoimmune Thyroid diseases. *J. Clin. Endocrinol. Metab.* **87**, 404–407 (2002).

34. Schuelke, M. An economic method for the fluorescent labeling of PCR fragments. *Nat. Biotechnol.* **18**, 233-234 (2000).

35. Surridge, A. K., Smith, A. C., Buchanan-Smith, H. M. & Mundy, N. I. Single-copy nuclear DNA sequences obtained from noninvasively collected primate feces. *Am. J. Primatol.* **56**, 185-90 (2002).

36. Taberlet, P. *et al.* Reliable genotyping of samples with very low DNA quantities using PCR. *Nucleic Acids Res.* **24**, 3189-94 (1996).

37. Van Oosterhout, C., Hutchinson, W. F., Wills, D. P. M. & Shipley, P. MICRO-CHECKER: Software for identifying and correcting genotyping errors in microsatellite data. *Mol. Ecol. Notes* **4**, 535-538 (2004).

38. Peakall, R. & Smouse, P. E. GenAlEx 6.5: genetic analysis in Excel. Population genetic software for teaching and research-an update. *Bioinformatics* **28**, 2537-2539 (2012).

39. Excoffier, L. & Lischer, H. E. L. ARLEQUIN suite ver 3.5: a new series of programs to perform population genetics analyses under Linux and Windows. *Mol. Ecol. Resour.* **10**, 564-567 (2010).

40. Goudet, J. Fstat (ver. 2.9.4), a program to estimate and test population genetics parameters. Available from http://www.unil.ch/izea/softwares/fstat.html Updated from Goudet 1995 (2003).

41. Pritchard, J. K., Stephens, M. & Donnelly, P. Inference of population structure using multilocus genotype data. *Genetics* **155**, 945-959 (2000).

42. Jakobsson, M. & Rosenberg, N. CLUMPP: A Cluster Matching and Permutation Program for Dealing with Label Switching and Multimodality in Analysis of Population Structure. *Bioinformatics* **23**, 1801–1806 (2007).

43. Rosenberg, N. A. Distruct: a program for the graphical display of population structure. *Mol. Ecol. Notes* **4**, 137-138 (2004).

44. Evanno, G., Regnaut, S. & Goudet, J. Detecting the number of clusters of individuals using the software STRUCTURE: a simulation study. *Mol. Ecol.* **14**, 2611-2620 (2005).

45. Piry, S. *et al.* GENECLASS2: A Software for Genetic Assignment and First-Generation Migrant Detection. *J. Heredity* **95**, 536–539 (2004).

46. Rannala, B. & Mountain, J. L. Detecting immigration by using multilocus genotypes. *Proc. Natl. Acad. Sci. U. S. A.* **94**, 5 (1997).

47. MAYDS Informe del Estado de Ambiente. 2016. Presidencia de la Nación, Argentina. Preprint at https://www.argentina.gob.ar/sites/default/files/mayds_informe_estado_ambiente_2016_baja_1_0.pdf (2016).

48. MAYDS Informe del Estado de Ambiente. 2017. Presidencia de la Nación, Argentina https://www.argentina.gob.ar/sites/default/files/completo-compressed.pdf (2017).

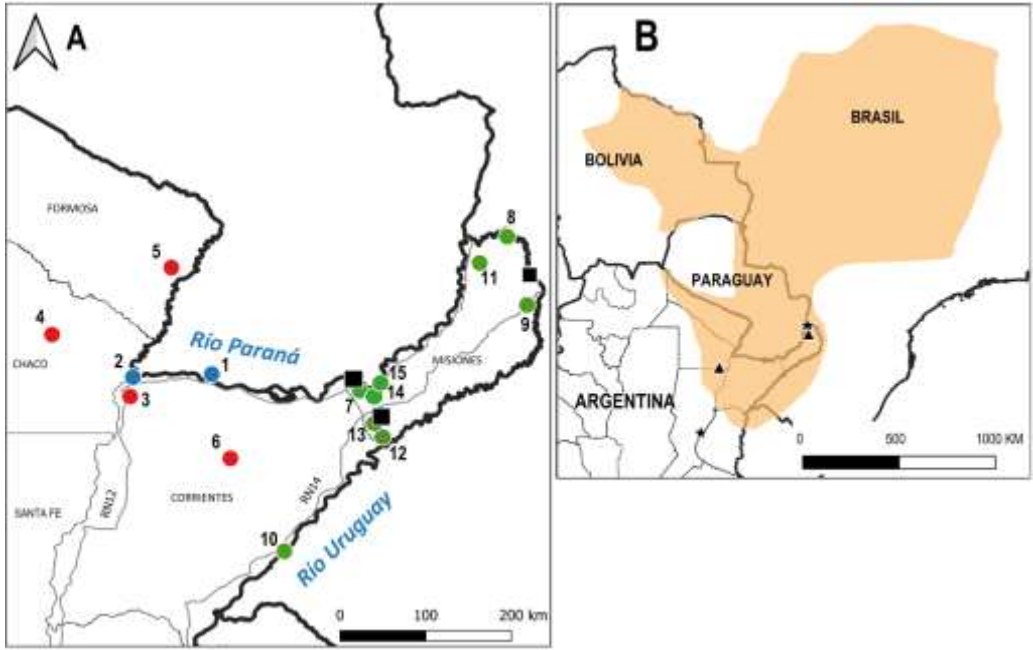

Figure 1: A - Map of the 15 populations included in the database of *A. caraya* genotypes in Argentina. Maps show (1 to 10) previously published and (11 to 15) newly sampled localities. Color-coded circles indicate the three genetic clusters identified using the structure analysis in the present study: blue: P-RP cluster, red: F-Ch-C cluster, and green: M-RU cluster. The complete names of sampling sites are listed in Table 1. Black squares indicate the sites were corpses of *A. caraya* were found. B - Map showing the distribution range of *A. caraya*. Black stars show the location of the rescue centers included in this study. Black triangles represent the reintroduction sites.

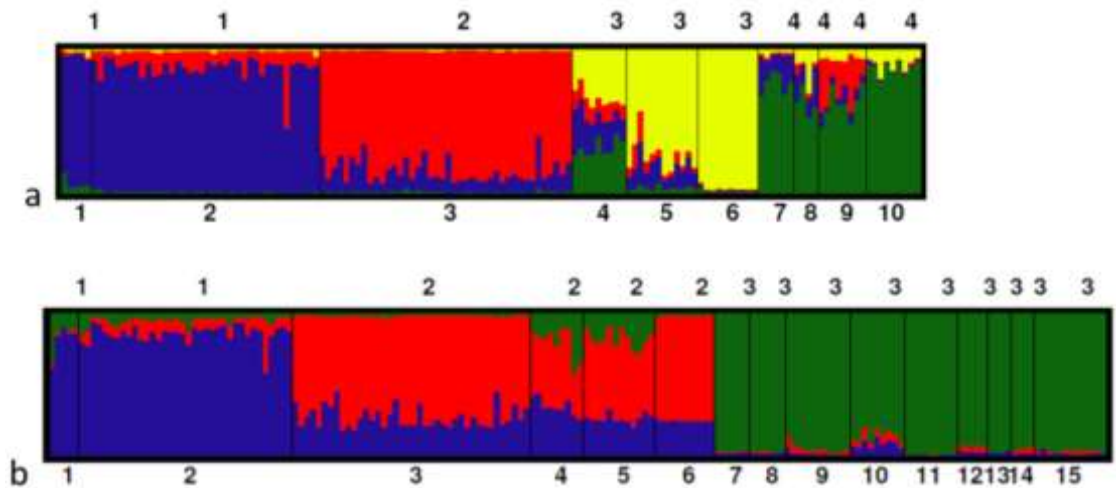

Figure 2: A) Structure analysis of clusters in *A. caraya*[8] (K = 4): blue: P-RP cluster 1, red: EBCO cluster 2, yellow: F-Ch-C cluster 3, and green: M-RU cluster 4. B) Structure analysis incorporating samples from the new localities (11 to 15) (K = 3) sampled in the present study: blue: P-RP cluster 1, red: F-Ch-C cluster 2, and green: M-RU cluster 3. Individuals are represented by vertical lines (y-axis) coloured in proportion to their membership coefficients in each cluster and grouped into populations of samples and separated with a black line. Complete names of populations are listed in Table 1.

Table 1: Total sampling for IGDB for *A. caraya* in Argentina. Populations sampled in Oklander[8] plus samples collected during the monitoring program for the epidemiological surveillance of YFV and other arboviroses in non-human primates in 2017 and 2018. Na = N° of different alleles; Ne = N° of effective alleles (calculated as $1/\Sigma(\text{allele frequency})^2$); He = expected heterozygosity = $1 - \Sigma(\text{allele frequency})^2$; uHe = unbiased expected heterozygosity = $(2N/(2N-1))*$ He; AR: allelic richness; PA = N° of alleles unique to a single population; FIS: inbreeding coefficient.

| Population number | Population name | Localization | | N samples | Na | | Ne | | He | | uHe | | AR | | PA | FIS |
|---|---|---|---|---|---|---|---|---|---|---|---|---|---|---|---|---|
| | | | | | Average | SD | Average | SD | Average | SD | Average | SD | Average | SD | | |
| 1 | Paraguay | -27.275 | -57.684 | 5.000 | 2.800 | 0.44 | 2.038 | 0.30 | 0.420 | 0.07 | 0.467 | 0.08 | 3.082 | 0.92 | 1 | 0.179 |
| 2 | Isla, Rio Paraná | -27.314 | -58.646 | 36.000 | 4.900 | 0.86 | 2.550 | 0.41 | 0.501 | 0.08 | 0.508 | 0.08 | 2.449 | 0.97 | 2 | 0.012 |
| 3 | EBCO, Corrientes | -27.550 | -58.679 | 40.000 | 4.500 | 0.93 | 2.165 | 0.34 | 0.440 | 0.08 | 0.445 | 0.07 | 2.619 | 1.09 | 2 | -0.039 |
| 4 | PN Chaco, Chaco | -26.794 | -59.618 | 9.000 | 2.900 | 0.28 | 1.813 | 0.15 | 0.410 | 0.05 | 0.435 | 0.05 | 2.779 | 1.01 | 1 | 0.084 |
| 5 | Guaycolec, Formosa | -25.985 | -58.161 | 12.000 | 3.700 | 0.54 | 2.285 | 0.27 | 0.495 | 0.06 | 0.517 | 0.07 | 2.211 | 1.09 | 1 | 0.080 |
| 6 | San Alonso, Corrientes | -28.306 | -57.456 | 10.000 | 2.700 | 0.47 | 1.914 | 0.31 | 0.356 | 0.09 | 0.374 | 0.09 | 2.500 | 0.71 | | -0.140 |
| 7 | Garupa, Misiones | -27.467 | -55.827 | 6.000 | 2.800 | 0.36 | 1.896 | 0.29 | 0.385 | 0.07 | 0.420 | 0.08 | 2.738 | 0.96 | | -0.169 |
| 8 | Yacutinga Lodge, Misiones | -25.574 | -54.075 | 6.000 | 2.400 | 0.43 | 1.877 | 0.31 | 0.376 | 0.07 | 0.411 | 0.08 | 2.200 | 0.79 | 1 | -0.198 |
| 9 | PP Piñalito, Misiones | -26.500 | -53.833 | 11.000 | 4.400 | 0.52 | 2.591 | 0.30 | 0.551 | 0.06 | 0.577 | 0.07 | 2.848 | 0.82 | 4 | 0.204 |
| 10 | Yapeyu, Corrientes | -29.445 | -56.800 | 9.000 | 3.600 | 0.30 | 2.414 | 0.30 | 0.509 | 0.07 | 0.539 | 0.07 | 2.784 | 0.82 | | -0.056 |
| 11 | PP Lago Urugua-í, Misiones | -25.921 | -54.419 | 9.000 | 3.500 | 0.52 | 2.359 | 0.33 | 0.504 | 0.07 | 0.533 | 0.07 | 2.178 | 1.03 | | 0.080 |
| 12 | Azara, Misiones | -27.984 | -55.787 | 5.000 | 2.500 | 0.27 | 1.841 | 0.19 | 0.394 | 0.07 | 0.438 | 0.07 | 2.773 | 1.12 | 1 | -0.271 |
| 13 | Apóstoles, Misiones | -27.910 | -55.761 | 4.000 | 2.500 | 0.22 | 1.865 | 0.12 | 0.441 | 0.04 | 0.504 | 0.05 | 2.593 | 1.18 | | -0.169 |
| 14 | Reserva Urutau EBY | -27.518 | -55.788 | 4.000 | 2.200 | 0.25 | 1.731 | 0.18 | 0.359 | 0.07 | 0.411 | 0.08 | 2.336 | 0.75 | | -0.113 |
| 15 | Sta Cecilia, Misiones | -27.429 | -55.710 | 12.000 | 3.800 | 0.51 | 2.330 | 0.22 | 0.522 | 0.06 | 0.545 | 0.06 | 2.347 | 0.55 | 2 | 0.085 |

Table 2: Genetic assignment of individuals using GeneClass2 and according to the criteria described by Rannala & Mountain[46] within the 15 populations that compose the database of genotypes for *A. caraya* in Argentina (column 1-4), and within the 3 genetic clusters identified for *A. caraya* in Argentina (column 5-8). Threshold: 0.05.

|  | Rank population | score | Rank population | score | Rank cluster | score | Rank cluster | score |
|---|---|---|---|---|---|---|---|---|
| Assigned sample | 1 | % | 2 | % | 1 | % | 2 | % |
| Güira-Oga 1 | Pop 6 | 96.46 | Pop 3 | 1.07 | F-Ch-C | 86.07 | P-RP | 13.94 |
| Güira-Oga 2 | Pop 5 | 84.56 | Pop 6 | 4.34 | F-Ch-C | 99.84 | P-RP | 0.16 |
| Güira-Oga 3 | Pop 3 | 44.28 | Pop 5 | 20.71 | F-Ch-C | 98.38 | M-RU | 1.62 |
| Güira-Oga 4 | Pop 2 | 75.75 | Pop 5 | 2.58 | P-RP | 93.51 | F-Ch-C | 6.49 |
| Güira-Oga 5 | Pop 7 | 38.42 | Pop 12 | 21.43 | M-RU | 96.57 | F-Ch-C | 3.26 |
| Güira-Oga 6 | Pop 3 | 39.59 | Pop 14 | 21.23 | F-Ch-C | 75.37 | P-RP | 24.35 |
| Güira-Oga 7 | Pop 3 | 49.60 | Pop 2 | 10.85 | F-Ch-C | 70.25 | P-RP | 29.59 |
| Güira-Oga 8 | Pop 3 | 98.06 | Pop 4 | 0.54 | F-Ch-C | 99.98 | P-RP | 0.02 |
| Güira-Oga 9 | Pop 5 | 70.05 | Pop 10 | 8.20 | F-Ch-C | 100.00 | M-RU | 0.00 |
| Güira-Oga 10 | Pop 2 | 98.99 | Pop 4 | 0.23 | P-RP | 99.59 | F-Ch-C | 0.41 |
| Güira-Oga 11 | Pop 2 | 44.47 | Pop 13 | 10.59 | P-RP | 99.31 | F-Ch-C | 0.59 |
| Güira-Oga 12 | Pop 3 | 49.41 | Pop 4 | 7.14 | F-Ch-C | 86.05 | P-RP | 12.97 |
| Güira-Oga 13 | Pop 3 | 96.65 | Pop 2 | 0.13 | F-Ch-C | 93.83 | P-RP | 6.15 |
| Güira-Oga 14 | Pop 3 | 99.81 | Pop 2 | 0.02 | F-Ch-C | 96.88 | P-RP | 3.10 |
| Güira-Oga 15 | Pop 3 | 96.54 | Pop 2 | 0.48 | F-Ch-C | 88.15 | P-RP | 11.80 |
| Güira-Oga 16 | Pop 13 | 74.15 | Pop 15 | 10.19 | M-RU | 99.97 | P-RP | 0.03 |
| Güira-Oga 17 | Pop 1 | 89.15 | Pop 2 | 0.06 | P-RP | 99.87 | F-Ch-C | 0.13 |
| Esmeralda 1 | Pop 10 | 53.12 | Pop 7 | 5.36 | M-RU | 99.37 | F-Ch-C | 0.61 |
| Esmeralda 2 | Pop 3 | 89.16 | Pop 2 | 0.44 | F-Ch-C | 81.00 | P-RP | 19.00 |
| Esmeralda 3 | Pop 3 | 62.74 | Pop 2 | 3.67 | F-Ch-C | 56.68 | P-RP | 43.32 |
| Esmeralda 4 | Pop 3 | 98.56 | Pop 2 | 0.10 | F-Ch-C | 97.92 | P-RP | 2.08 |
| Esmeralda 5 | Pop 6 | 74.28 | Pop 1 | 4.05 | F-Ch-C | 99.15 | P-RP | 0.85 |
| Found dead in Pop 13, Misiones | Pop 5 | 80.29 | Pop 9 | 2.08 | F-Ch-C | 98.04 | M-RU | 1.49 |
| Found dead in Posadas, Misiones | Pop 15 | 54.69 | Pop 13 | 14.29 | M-RU | 69.54 | P-RP | 25.74 |
| Found dead in San Antonio, Misiones | Pop 12 | 42.38 | Pop 15 | 19.37 | M-RU | 99.67 | F-Ch-C | 0.23 |